\documentclass[twocolumn,paper]{ieice}
\usepackage[dvipdf]{graphicx}
\usepackage{latexsym}
\usepackage{amsmath}
\usepackage{amssymb}
\usepackage{color}
\usepackage{bm}
\usepackage{amsmath}
\usepackage{amssymb}
\usepackage{amsfonts}
\usepackage{amsthm}
\usepackage{bm}
\usepackage{amsmath}
\usepackage{bm}	% required for `\bm' (yatex added)
\usepackage{amssymb}
\usepackage{bm}

\newcommand{\Vt}{V_{\mathrm{t}}}
\newcommand{\Vp}{V_{\mathrm{p}}}

\newcommand{\epsBP}{\epsilon_{\mathrm{BP}}}
\newcommand{\epsMAP}{\epsilon_{\mathrm{MAP}}}
\newcommand{\xs}{\bm{s}}
\newcommand{\xn}{\bm{n}}
\newcommand{\Cs}{H_1}
\newcommand{\Cn}{H_2}
\newcommand{\Ct}{H_3}
\newcommand{\Cm}{H_4}

\newcommand{\dk}{{\mathtt{k}}}
\newcommand{\dr}{{\mathtt{r}}}
\newcommand{\dg}{{\mathtt{g}}}
\newcommand{\dl}{{\mathtt{l}}}
\newcommand{\hL}{\hat{L}}

\vol{xx}
\no{xx}
\field{A}
\SpecialSection{Information Theory and Its Applications}
\title{Spatially-Coupled MacKay-Neal Codes and Hsu-Anastasopoulos Codes}
\authorlist{
 \authorentry[kenta@comm.ss.titech.ac.jp]{Kenta KASAI}{m}{TITECH}
 \authorentry[sakaniwa@ss.titech.ac.jp]{Kohichi SAKANIWA}{f}{TITECH}
}
\affiliate[TITECH]{Dept.\ of Communications and Integrated Systems, 
Tokyo Institute of Technology
2-12-1, O-Okayama, Meguro-ku, Tokyo, 152-8550, Japan}
\usepackage{color}	% required for `\color' (yatex added)
\begin{document}
\maketitle
\begin{summary}
Kudekar {\it et al.} recently proved that for transmission over the binary erasure channel (BEC), 
spatial coupling of LDPC codes increases the BP threshold of the coupled ensemble to 
the MAP threshold of the underlying LDPC codes. One major drawback of the capacity-achieving 
spatially-coupled LDPC codes is that one needs to increase the column and row weight of 
parity-check matrices of the underlying LDPC codes.

It is proved, that Hsu-Anastasopoulos (HA) codes and MacKay-Neal (MN) codes
achieve the capacity of memoryless binary-input symmetric-output channels under 
MAP decoding with bounded column and row weight of the parity-check matrices. 
The HA codes and the MN codes are dual codes each other.

The aim of this paper is to present an empirical evidence that spatially-coupled MN (resp.~HA) codes with bounded column and row weight 
achieve the capacity of the BEC. 
To this end, we introduce a spatial coupling scheme of MN (resp.~HA) codes.
By density evolution analysis, we will show that the resulting spatially-coupled
MN (resp.~HA) codes have the BP threshold close to the Shannon limit.
\end{summary}

%%%%%%%%%%%%%%%%%%%%%%%%%%%%%%%%%%%%%%%%%%%%%%%%%%%%%%%%%%%%%%%
\section{Introduction}
%%%%%%%%%%%%%%%%%%%%%%%%%%%%%%%%%%%%%%%%%%%%%%%%%%%%%%%%%%%%%%%
Achieving the capacity of memoryless binary-input symmetric-output (MBS) channels \cite[Chap.~4]{mct} under efficient decoding algorithms used to be an ultimate goal for coding theorists. 
The capacity of MBS channels was practically achieved by irregular LDPC codes 
\cite{mct} with belief propagation (BP) decoding and rigorously achieved by polar codes \cite{5075875} with successive cancellation.
Spatially-coupled (SC) LDPC codes \cite{5695130} have bounded range of parity-check constrains like convolutional codes. 
Recently, SC-LDPC codes have attracted much attention due to the fact that the codes achieves the capacity of binary erasure channels (BEC) \cite{5695130} and 
an observation that the codes seem to achieve the capacity of MBS channels \cite{DBLP:journals/corr/abs-1004-3742}.

SC-LDPC codes are capacity-achieving codes designed based on the construction of convolutional LDPC codes.
Felstr{\"o}m and Zigangirov \cite{zigangirov99} introduced a construction method of $(\dl, \dr)$-regular convolutional LDPC codes from $(\dl, \dr)$-regular block LDPC codes \cite{mct}. 
The convolutional LDPC codes exhibited better decoding performance than the underlying block LDPC codes under a fair comparison with respect to the code length. 
Note that in this paper,  convolutional LDPC codes are defined by sparse band parity-check matrices. 
Lentmaier {\it et al.}~ observed that (4,8)-regular convolutional LDPC codes exhibited the decoding performance surpassing the BP threshold of (4,8)-regular block LDPC codes \cite{lentmaier_II}. 
Sridharan {\it et al.}~ developed the density evolution (DE) \cite{mct} for the BEC and calculated the BP threshold \cite{Sridharan_Allerton2004}. 
Lentmaier {\it et al.}~ developed the DE for the binary-input memoryless (BMS) channels \cite{5571910}. 
Further, the BP threshold equals to the MAP threshold of the underlying block LDPC codes with a lot of accuracy. 
The MAP threshold  was calculated by the extended BP (EBP) extrinsic information transfer (EXIT) function analysis \cite{5290273}. 
%This fact is not coincidence. 
Constructing convolutional LDPC codes from a block LDPC code improves the BP threshold up to the MAP threshold of the underlying codes. 

Kudekar {\it et al.}~ named this phenomenon  ``threshold saturation'' and proved rigorously for the BEC \cite{5695130}. 
In the limit of large $L$ and $w$, 
the SC-LDPC code ensemble $(\dl,\dr,L,w)$ \cite{5695130} was shown to achieve the MAP threshold of $(\dl,\dr)$-regular LDPC code ensemble. 
The parameters $L$ and $w$ are the coupling number and the randomized window size. 
For more details, we refer the readers to \cite{5695130}.

Further, by computing EBP generalized EXIT (GEXIT) curves \cite{5290273}, Kudekar {\it et al.}~ \cite{DBLP:journals/corr/abs-1004-3742} 
observed empirical evidence which supports the threshold saturation occurs also for the BMS channels.
%The design coding rate of $(\dl, \dr)$-regular block LDPC codes is $1-\dl/\dr$. 
For arbitrary BMS channels, the MAP threshold of the codes quickly converge to the Shannon limit while the BP threshold goes to 0 \cite{959254}. 
%For example, in Fig.~\ref{234331_29Sep10} we listed the BP and MAP threshold of $(\dl,\dr)$-regular block LDPC codes of rate 1/2.
In other words, in the limit of large $\dl$ as keeping $\frac{\dl}{\dr}$, $L$ and $w$,  the SC-LDPC code ensemble $(\dl,\dr,L,w)$ 
achieves {\em universally} the capacity of the BMS channels under BP decoding. 
Such universality is not supported by other efficiently-decodable capacity-achieving codes, i.e., 
polar codes \cite{5075875} and irregular LDPC codes \cite{richardson01design}.
According to the channel, polar codes need selection of frozen bits \cite{DBLP:journals/corr/abs-0901-2207}
and irregular LDPC codes need optimization of degree distributions.

% \begin{table}[t]
% \caption{The BP and MAP thresholds of $(\dl,\dr)$-regular block LDPC \cite{5290273}}
% \label{234331_29Sep10}
%  \begin{center}
%    \begin{tabular}{ccc}
%  $(\dl,\dr)$ & $\epsilon_{\mathrm{BP}}$ & $\epsilon_{\mathrm{MAP}}$ \\\hline
%  (3,6) & 0.4294&0.48815\\
%  (4,8) & 0.3834&0.49774\\
%  (5,10)& 0.3416&0.49949\\
%  (6,12)& 0.3075&0.49988\\
%  (7,14)& 0.2797&0.49997
%  \end{tabular}
% \end{center}
% \end{table}

One major drawback of the $(\dl,\dr,L,w)$ code ensemble is that one needs to increase degree $\dl$ and $\dr$ to strictly achieve the capacity. 
The number of non-zero entries in the parity-check matrices is proportional to the decoding computations of BP decoding. 
Specifically, for the BEC case, this is identical to the total computations for decoding \cite{935864}. 
The number of edges per information bit is called {\em density}. 
For example the density of $(\dl,\dr)$-regular LDPC codes is $\frac{\dl\dr}{\dr-\dl}$. 
For fixed $R=1-\frac{\dl}{\dr}$, the density is unbounded as $\dl,\dr\to\infty$. 
It is desired to reduce the density. 
Lentmaier {\it et al.} constructed coupled Accumulate-Repeat-Jagged-Accumulate (ARJA) codes \cite{5513633} which exhibit very good BP threshold. 
The density of the ARJA code is bounded but the BP threshold leaves a small gap to the Shannon limit. 

For any code achieving a fraction $1-\epsilon$ of capacity of the BMS under MAP decoding, 
Sason and Urbanke showed that the density of the code is at least $\frac{K_1+\ln \frac{1}{\epsilon}}{K_2}$, where $K_1$ and $K_2$ are constant \cite{1207364}. 
In other words, the density of capacity-achieving codes needs to be unbounded. 
However, this is not the case for the codes with puncture. 
The punctured bits work as auxiliary states and help decoding. 
Such puncturing is widely used for structured codes \cite{met}. 
Pfister and Sason constructed Accumulate-Repeat-Accumulate (ARA) codes and Accumulate-LDPC (ALDPC) codes which achieve the capacity of BEC with bounded density \cite{4215147}. 

MacKay-Neal codes \cite{mn_code} are non-systematic two-edge type LDPC codes \cite{met,mct}. The MacKay-Neal (MN) codes are conjectured to achieve the capacity of BMS channels under ML decoding. 
Murayama {\it et al.}~\cite{PhysRevE.62.1577} and Tanaka {\it et al.}~\cite{TanakaSaad2003} reported the empirical evidence of the conjecture for BSC and AWGN channels, respectively 
by a non-rigorous statistical mechanics approach known as {\em replica method}. 

Hsu and Anastasopoulos \cite{5429143} rigorously proved that LDPC codes concatenated with LDGM (low-density generator-matrix) codes \cite{ldgm} achieve the capacity of arbitrary BMS channels with bounded density under ML decoding. 
We name the codes  Hsu-Anastasopoulos (HA) codes after the inventors.
Furthermore, Wainwright and Martinian showed HA codes achieve the rate-distortion bound for symmetric Bernoulli sources \cite{4787613}. 

On the contrary to their ML decoding performance, it is interesting to see that the MN and HA codes have no BP thresholds. 
In other words, the bit error rate under BP decoding does not go to 0, even if none of the transmitted bits are erased.
%In other words, the bit error rate does not vanish in the limit of large code length even when the all transmitted bits are known to the receiver.
%On the other hand, the MAP thresholds of the MN and HA codes are equal to the Shannon limit.

The aim of this paper is to present an empirical evidence that SC-MN (resp.~SC-HA) codes with bounded density 
achieve the capacity of the BEC. 
To this end, we introduce a spatial coupling scheme of MN (resp.~HA) codes. 
Spatial coupling of MN and HA codes has never been studied before. 
By DE analysis, we will show that the resulting SC-MN (resp.~SC-HA) codes have the BP threshold close to the Shannon limit.

\begin{figure}[t!]
 \begin{center}
 \includegraphics[scale=0.5]{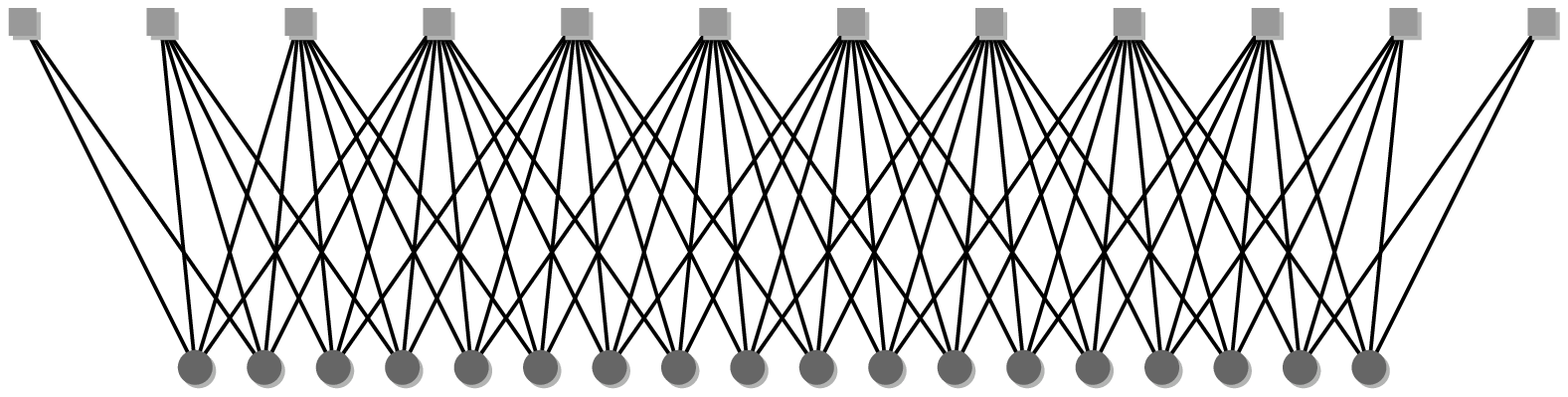}
 \end{center}
\caption{Protograph of $ (4, 8, 4)$ SC-LDPC codes}
\label{214346_29Sep10}
\end{figure}

% The aim of this paper is to construct bounded-density codes achieving the capacity of arbitrary BMS channels. 
% To this end, we consider spatial coupling of MN codes and HA codes. 
% The rest of this paper is organized as follows. 
% Section \ref{191301_30Sep10} defines MN codes and the density evolution (DE) and calculates fixed points (FP) of the DE.
% Section \ref{191403_30Sep10} introduces spatially-coupled MN codes and their DE and calcutes the FP the DE and the BP threshold. 
% Section \ref{232945_30Sep10} concludes this paper.
%%%%%%%%%%%%%%%%%%%%%%%%%%%%%%%%%%%%%%%%%%%%
\section{Preliminaries}
%%%%%%%%%%%%%%%%%%%%%%%%%%%%%%%%%%%%%%%%%%%%
In this section, we briefly review the $(\dl,\dr, L)$ SC-LDPC codes introduced by Kudekar {\it et al.}~\cite{5695130} 
and their results of performance analysis. We assume $\frac{\dr}{\dl}=:\dk\in\mathbb{Z}.$
For simplicity, we focus on rate 1/2 codes, i.e.,  $\dk=2$.

The SC-LDPC codes are defined by the following protograph codes \cite{protograph}. 
The adjacency matrix of the protograph is referred to as a base matrix. 
The base matrix of ($\dl,\dr,L$) SC-LDPC code is given as follow. 
Let us define as $\hL:=2L+1$.
Let $H(\dl, \dr, \hL,\dk)$ be an $(\hL+\dl-1)\times \dk\hL$ band binary matrix 
of band size $\dr\times\dl$ and column weight $\dl$, where the band size is height $\times$ width of the band.
For example $H(4,8,9,2)$ is given in Fig.~\ref{045049_15Feb11}.
The $(\dl,\dr,L)$ SC-LDPC codes is defined as protograph codes defined by base matrix $H(\dl, \dr, \hL,\dk)$. 
Figure \ref{214346_29Sep10} shows protograph of $(4, 8, 4)$ SC-LDPC codes.
The protograph of $(\dl, \dr, L)$ SC-LDPC codes have $\dk\hL$ variable nodes and $\hL+\dl-1$ check nodes.
Hence, the design coding rate of $(\dl,\dr,L)$ SC-LDPC codes is given by
\begin{align*}
 R(\dl, \dr, L)&=\frac{\dk\hL-(\hL+\dl-1)}{\dk\hL}=\frac{\dk-1}{\dk}-\frac{\dl-1}{\dk\hL}, 
\end{align*}
which converges to $(\dk-1)/\dk$ as $\hL\to\infty$ with vanishing gap like $O(1/\hL).$ 

Table \ref{144758_30Sep10} shows BP threshold values $\epsBP(\dl,\dr,L)$ of $(\dl,\dr,L)$ SC-LDPC codes. 
As $L$ increases, it is observed that the design coding rate $R(3,6,L)$ converges to 1/2 and the BP threshold values $\epsBP$ 
approach the MAP threshold value of $(3,6)$ LDPC codes $\epsMAP(\dl,\dr)\approx 0.488151$. 

Kudekar {\it et al.} observed $\epsBP(\dl,\dr,L)$ does not converge to $\epsMAP(\dl,\dr)$. There remains a small gap of order $10^{-6}$ in significant digits. 
In order to decrease the gap, Kudekar {\it et al.} introduced the $(\dl,\dr,L,w)$ SC-LDPC codes 
that allows randomized connection of edges of window size $w$ \cite{5695130}. 
It is shown that $\epsBP(\dl,\dr,L,w)$ converges to $\epsMAP(\dl,\dr)$ as $L$ and $w$ tend to infinity as 
\begin{align*}
 \lim_{w\to\infty}\lim_{L\to\infty}\epsMAP(\dl,\dr,L,w)=\epsMAP(\dl,\dr).
\end{align*}
It is known \cite{5695130} that $\epsMAP(\dl,\dr)$ quickly converges to the Shannon limit $1-R$, i.e., 
\begin{align*}
 \lim_{\substack{R=1-\dl/\dr\\\dl\to\infty}}\epsMAP(\dl,\dr)=1-R.
\end{align*}
Table \ref{234331_29Sep10} shows the MAP threshold values of $(\dl,\dr)$ LDPC codes. 
\begin{table}[t]
\caption{The BP and MAP threshold values of $(\dl,\dr)$ LDPC codes.
The MAP threshold  $\epsMAP(\dl,\dr)$ quickly converges to the Shannon limit $1-R$.
While the BP threshold  converges to 0.}
\label{234331_29Sep10}
 \begin{center}
 \begin{tabular}{ccc}
 $(\dl,\dr)$ & $\epsilon_{\mathrm{BP}}$ & $\epsilon_{\mathrm{MAP}}$ \\\hline
 (3,6) & 0.4294&0.48815\\
 (4,8) & 0.3834&0.49774\\
 (5,10)& 0.3416&0.49949\\
 (6,12)& 0.3075&0.49988\\
 (7,14)& 0.2797&0.49997
 \end{tabular}
 \end{center}
\end{table}
This implies that SC-LDPC codes achieve the capacity of the BEC, in the limit of large column and row weight. 
On the other hand, infinite column and row weight are required for the codes to achieve the capacity. 

\begin{table}[t]
\caption{BP threshold values $\epsBP$ of $(3,6,L)$ SC-LDPC and design coding rate $R$ \cite{5695130}}
\label{144758_30Sep10}
\begin{center}
  \begin{tabular}{ccc}
 $L$& $\epsBP$ &$R$\\\hline
   1 &0.714309 &0.166667\\
   2 &0.587842 &0.300000\\
   4 &0.512034 &0.388889\\
   8 &0.488757 &0.441176\\
  16 &0.488151 &0.469697\\
  32 &0.488151 &0.484615\\
  64 &0.488151 &0.492248\\
 128 &0.488151 &0.496109
 \end{tabular}
\end{center}
\end{table}

%%%%%%%%%%%%%%%%%%%%%%%%%%%%%%%%%%%%%%%%%%%%
\section{Bounded-Density Codes Achieving Capacity under ML Decoding}
%%%%%%%%%%%%%%%%%%%%%%%%%%%%%%%%%%%%%%%%%%%%
In this section, we give the definition of  MN codes and HA codes. 
%%%%%%%%%%%%%%%%%%%%%%%%%%%%%%%%%%%%%%%%%%%%
\subsection{MacKay-Neal codes}
\label{191301_30Sep10}
%%%%%%%%%%%%%%%%%%%%%%%%%%%%%%%%%%%%%%%%%%%%
Let $\Cs$ be a random binary matrix of size $N\times\frac{\dr}{\dl}N$ with column weight $\dl$ and row weight $\dr$. 
Let $\Cn$ be a random binary matrix of size $N\times N$ with column weight $\dg$ and row weight $\dg$. 
\begin{figure*}[t]
\begin{center}
 \renewcommand\arraystretch{0.5}
 $
 H(4,8,9,2)=
  \left[
  \begin{minipage}{2.7cm}
 \begin{tabular}{@{}l@{}l@{}l@{}l@{}l@{}l@{}l@{}l@{}l@{}l@{}l@{}l@{}l@{}l@{}l@{}l@{}l@{}l}
 1&1& & & & & & & & & & & & & & & & \\
 1&1&1&1& & & & & & & & & & & & & & \\
 1&1&1&1&1&1& & & & & & & & & & & & \\
 1&1&1&1&1&1&1&1& & & & & & & & & & \\
 & &1&1&1&1&1&1&1&1& & & & & & & & \\
 & & & &1&1&1&1&1&1&1&1& & & & & & \\
 & & & & & &1&1&1&1&1&1&1&1& & & & \\
 & & & & & & & &1&1&1&1&1&1&1&1& & \\
 & & & & & & & & & &1&1&1&1&1&1&1&1\\
 & & & & & & & & & & & &1&1&1&1&1&1\\
 & & & & & & & & & & & & & &1&1&1&1\\
 & & & & & & & & & & & & & & & &1&1
 \end{tabular}
 \end{minipage}
 \right],
 S(5,18)=
  \left[
  \begin{minipage}{2.7cm}
 \begin{tabular}{@{}l@{}l@{}l@{}l@{}l@{}l@{}l@{}l@{}l@{}l@{}l@{}l@{}l@{}l@{}l@{}l@{}l@{}l}
 1& & & & & & & & & & & & & & & & & \\
 1&1& & & & & & & & & & & & & & & & \\
 1&1&1& & & & & & & & & & & & & & & \\
 1&1&1&1& & & & & & & & & & & & & & \\
 1&1&1&1&1& & & & & & & & & & & & & \\
 &1&1&1&1&1& & & & & & & & & & & & \\
 & &1&1&1&1&1& & & & & & & & & & & \\
 & & &1&1&1&1&1& & & & & & & & & & \\
 & & & &1&1&1&1&1& & & & & & & & & \\
 & & & & &1&1&1&1&1& & & & & & & & \\
 & & & & & &1&1&1&1&1& & & & & & & \\
 & & & & & & &1&1&1&1&1& & & & & & \\
 & & & & & & & &1&1&1&1&1& & & & & \\
 & & & & & & & & &1&1&1&1&1& & & & \\
 & & & & & & & & & &1&1&1&1&1& & & \\
 & & & & & & & & & & &1&1&1&1&1& & \\
 & & & & & & & & & & & &1&1&1&1&1& \\
 & & & & & & & & & & & & &1&1&1&1&1\\
 & & & & & & & & & & & & & &1&1&1&1\\
 & & & & & & & & & & & & & & &1&1&1\\
 & & & & & & & & & & & & & & & &1&1\\
 & & & & & & & & & & & & & & & & &1
 \end{tabular}
 \end{minipage}
 \right]
 ,
 V(8,4,8,2)=\left[
  \begin{minipage}{1.2cm}
 \begin{tabular}{@{}l@{}l@{}l@{}l@{}l@{}l@{}l@{}l}
 1& & & & & & & \\
 1& & & & & & & \\
 1&1& & & & & & \\
 1&1& & & & & & \\
 1&1&1& & & & & \\
 1&1&1& & & & & \\
 1&1&1&1& & & & \\
 1&1&1&1& & & & \\
 &1&1&1&1& & & \\
 &1&1&1&1& & & \\
 & &1&1&1&1& & \\
 & &1&1&1&1& & \\
 & & &1&1&1&1& \\
 & & &1&1&1&1& \\
 & & & &1&1&1&1\\
 & & & &1&1&1&1\\
 & & & & &1&1&1\\
 & & & & &1&1&1\\
 & & & & & &1&1\\
 & & & & & &1&1\\
 & & & & & & &1\\
 & & & & & & &1
 \end{tabular}
 \end{minipage}
 \right]
 $
\end{center}
\renewcommand\arraystretch{1.0}

\caption{Matrices used for definition of SC-MN and SC-HA codes. 
For example,  $H(\dl=4, \dr=8, \hL=9,\dk=2)$ is a $(9+4-1)\times 2\cdot 9$ band binary matrix of band size $8\times 4$ 
and column weight $2$, where the band size is height $\times$ width of the band.
}
\label{045049_15Feb11}
\end{figure*}
An $(\dl,\dr,\dg)$-MN code is defined as an LDPC code with parity-check matrix
\begin{align}
 (\Cs\  \Cn)\label{072942_15Feb11} 
\end{align}
with $\dl,\dr,\dg\ge 2$ and such that the bits corresponding to $\Cs$ are punctured by the non-systematic fashion of the codes. 
Sparse parity-check representation is given by $\Cs\xs +\Cn\xn=0$, with information bits $\xs$ as state bits and parity bits $\xn$.
From $(\Cn)^{-1}\Cs\xs =\xn$, it follows that  
the generator matrix of the $(\dl,\dr,\dg)$-MN code is given by 
\begin{align}
G_{\mathrm{MN}}= \Cs^{\mathsf{T}}(\Cn^{-1})^{\mathsf{T}} \in \{0,1\}^{\frac{\dr}{\dl}N\times N}.\label{150950_10Feb11}
\end{align}
%Using this, MN codes are efficiently decoded by iterative decoding. 
The MN codes are non-systematic, in other words, only the parity bits $\xn$ are transmitted through the channel. 

The $(\dl,\dr,2)$-MN codes are called non-systematic Repeat-Accumulate (RA)  codes in \cite{4215147}.
The $(\dl,\dr,1)$-MN codes are identical to  ($\dl,\dr$)-LDGM  codes.

We give another definition of MN codes. Both definitions are equivalent in terms of DE.
The MN codes can be  defined by a multi-edge type LDPC code ensemble \cite{met} with degree distribution pair 
\begin{align*}
 \nu(\epsilon,x_1,x_2)&=\frac{\dr}{\dl}x_1^{\dl}+\epsilon x_2^{\dg},\\
 \mu(x_1,x_2)&=x_1^{\dr}x_2^{\dg}. 
\end{align*}
The design coding rate of an $(\dl,\dr,\dg)$-MN code is given by $\frac{\dr}{\dl}$.

Murayama {\it et al.}~\cite{PhysRevE.62.1577} and Tanaka {\it et al.}~\cite{TanakaSaad2003} reported empirical evidences that 
MN codes achieve the capacity of BSC and AWGN channels under ML decoding, respectively 
by a non-rigorous statistical mechanics approach known as {\em replica method}. 
From those results, we expect that $(\dl,\dr,\dg)$ MN codes achieve the capacity of arbitrary MBS channel 
if respectively $\dl,\dr,\dg\ge 2$ under ML decoding. 
Hence, we will use $\dl,\dr,\dg\ge 2$. 

Let $x^{(\ell)}$ and $y^{(\ell)}$ be the erasure probability of messages from information and parity bit nodes in the $\ell$-th round of BP decoding, respectively. 
DE gives the update equations of density $x^{(\ell)}$ and $y^{(\ell)}$ as follows. 
\begin{align*}
   x^{(\ell+1)}&=(1-(1-x^{(\ell)})^{\dr-1}(1-y^{(\ell)})^{\dg})^{\dl-1},\\
 y^{(\ell+1)}&=\epsilon (1-(1-x^{(\ell)})^{\dr}(1-y^{(\ell)})^{\dg-1})^{\dg-1},\\
 x^{(0)}&=1,\quad   y^{(0)}=1. 
\end{align*}
It is obvious that  $x^{(\ell)}=1, y^{(\ell)}=\epsilon$ for $\ell\ge 1$ for any $\epsilon\ge 0$. 
It follows $x^{(\ell)}$ does not converge to 0 as $\ell\to\infty$, even if $\epsilon=0$. 
Hence, MN codes have no BP threshold. 
%%%%%%%%%%%%%%%%%%%%%%%%%%%%%%%%%%%%%%%%%%%%%%%%%%%%
\subsection{Hsu-Anastasopoulos Codes}
%%%%%%%%%%%%%%%%%%%%%%%%%%%%%%%%%%%%%%%%%%%%%%%%%%%%
An $(\dl',\dr',\dg)$-HA code is a concatenation of 
an $(\dl',\dr')$-LDPC code and a $(\dg,\dg)$-LDGM code. 
Let $\Ct^{\mathsf{T}}$ be a random binary matrix of size $\frac{\dl'}{\dr'}N\times N$ with column weight $\dl'$ and row weight $\dr'$. 
Let $\Cm^{\mathsf{T}}$ be a random binary matrix of size $N\times N$ with column weight $\dg$ and row weight $\dg$. 
An $(\dl,\dr,\dg)$-HA code is defined as an LDPC code with parity-check matrix
\begin{align}
 \begin{pmatrix}
 \Ct^{\mathsf{T}} &O\\
 \Cm^{\mathsf{T}} &I
 \end{pmatrix}
\end{align}
with $\dl,\dr,\dg\ge 2$ and such that the bits corresponding to $\Ct^{\mathsf{T}}$ and $\Cm^{\mathsf{T}}$ are punctured. 
It follows that the parity-check matrix of the  $(\dl',\dr',\dg)$-HA code is given by
\begin{align}
 H_{\mathrm{HA}}=\Ct^{\mathsf{T}}(\Cm^{\mathsf{T}})^{-1}.\label{150955_10Feb11} 
\end{align}

From Eqs.~\eqref{150950_10Feb11} and \eqref{150955_10Feb11} it follows that 
if we set $\dl'=\dr$, $\dr'=\dl$, $\Cs=\Ct$ and $\Cn=\Cm$, it holds $G_{\mathrm{MN}}=H_{\mathrm{HA}}$.
It follows the $(\dl',\dr',\dg)$-HA code is dual code of the $(\dl,\dr,\dg)$-MN code.

The HA codes can be defined by a multi-edge type LDPC code ensemble \cite{met} with degree distribution pair 
\begin{align*}
 \nu(\epsilon,x_1,x_2,x_3)&=x_1^{\dl}x_2^{\dg}+\epsilon x_3,\\
 \mu(x_1,x_2)&=\frac{\dl}{\dr}x_1^{\dr} + x_2^{\dg}x_3. 
\end{align*}
The design coding rate of an $(\dl,\dr,\dg)$-HA code is given by $1-\frac{\dl}{\dr}$.
The $(\dl,\dr,2)$-HA code is referred to an accumulate LDPC code in \cite{4215147}.
The $(\dl,\dr,1)$-HA code is identical to an $(\dl,\dr)$-regular LDPC code.
Sparse parity-check representation is given by $\Cs\xs=0$ and $\Cn\xs=\xn$, with state bits $\xs$ and parity bits $\xn$.

For the arbitrary MBS channels,  it is shown that $(\dl,\dr,\dg)$-HA codes achieve the capacity under ML decoding with bounded $\dl,\dr,\dg$ \cite{5429143}.
% %%%%%%%%%%%%%%%%%%%%%%%%%%%%%%%%%%%%%%%%%%%%%%%%%%%%
% \section{Accumulate-Repeat-Accumulate Codes}
% %%%%%%%%%%%%%%%%%%%%%%%%%%%%%%%%%%%%%%%%%%%%%%%%%%%%
% \begin{align*}
%  W&=1-(1-x)^2\\
%  x&=Y^{\dl}X\\
%  X&=1-(1-\epsilon)(1-x)\\
%  y&=Y^{\dl-1}X^2\\
%  Y&=1-(1-y)^{\dr-1}(1-z)^2\\
%  z&=\epsilon (1-(1-y)^{\dr}(1-z))\\
%  z&= \epsilon\frac{1-(1-y)^{\dr}}{1-\epsilon(1-y)^{\dr}}\\
%  h(x,y,z)&=(\dr W + \dl Z^2)/(\dl+\dr)
% \end{align*}
%%%%%%%%%%%%%%%%%%%%%%%%%%%%%%%%%%%%%%%%%%%%%%%%%%%%
\section{Spatially-Coupled MN and HA Codes}
%%%%%%%%%%%%%%%%%%%%%%%%%%%%%%%%%%%%%%%%%%%%%%%%%%%%
%%%%%%%%%%%%%%%%%%%%%%%%%%%%%%%%%%%%%%%%%%%%%%%%%%%%
\subsection{Spatially-Coupled MacKay-Neal Codes}
%%%%%%%%%%%%%%%%%%%%%%%%%%%%%%%%%%%%%%%%%%%%%%%%%%%%
In this section, we propose spatial coupling of MN codes and evaluate the BP threshold values. 
For simplicity we focus on MN codes with $\dl=\dk\dr$, where $\dk\in\mathbb{Z}$.
We define SC-MN codes as follows. 
First, let $S(\dg, W)$ be a $(W+\dg-1)\times W$ binary band matrix 
of band size $\dg\times \dg$ and column weight $\dg$.
For example $S(5,18)$ is given in Fig.~\ref{045049_15Feb11}.
Next, let $V(\dl,\dr,\hL,\dk)$ be a binary band $(\dk\hL+\dl-2)\times \hL$ matrix of band size $\dl\times \dr$ and column weight $\dl$.
For example, $V(8,4,8,2)$ is given in Fig.~\ref{045049_15Feb11}. 
An $(\dl,\dr,\dg,L)$ SC-MN code is defined as a protograph code which is defined by base matrix
\begin{align}
 &\begin{pmatrix}
   V(\dl,\dr,\hL,\dk)&S(\dg,\dk\hL+\dl-\dg-1)
  \end{pmatrix}\label{072352_15Feb11}
\end{align}
and such that the bits corresponding to $V(\dl,\dr,\hL,\dk)$ are punctured.
Equations \eqref{072942_15Feb11} and \eqref{072352_15Feb11} are analogous in such a way that 
both have the same column-weight and almost the same row-weight distributions, 
and one is sparse matrix and the other is sparse band matrix under some permutation of  rows and columns.

\begin{figure}[t]
 \begin{center}
 \includegraphics[scale=0.5]{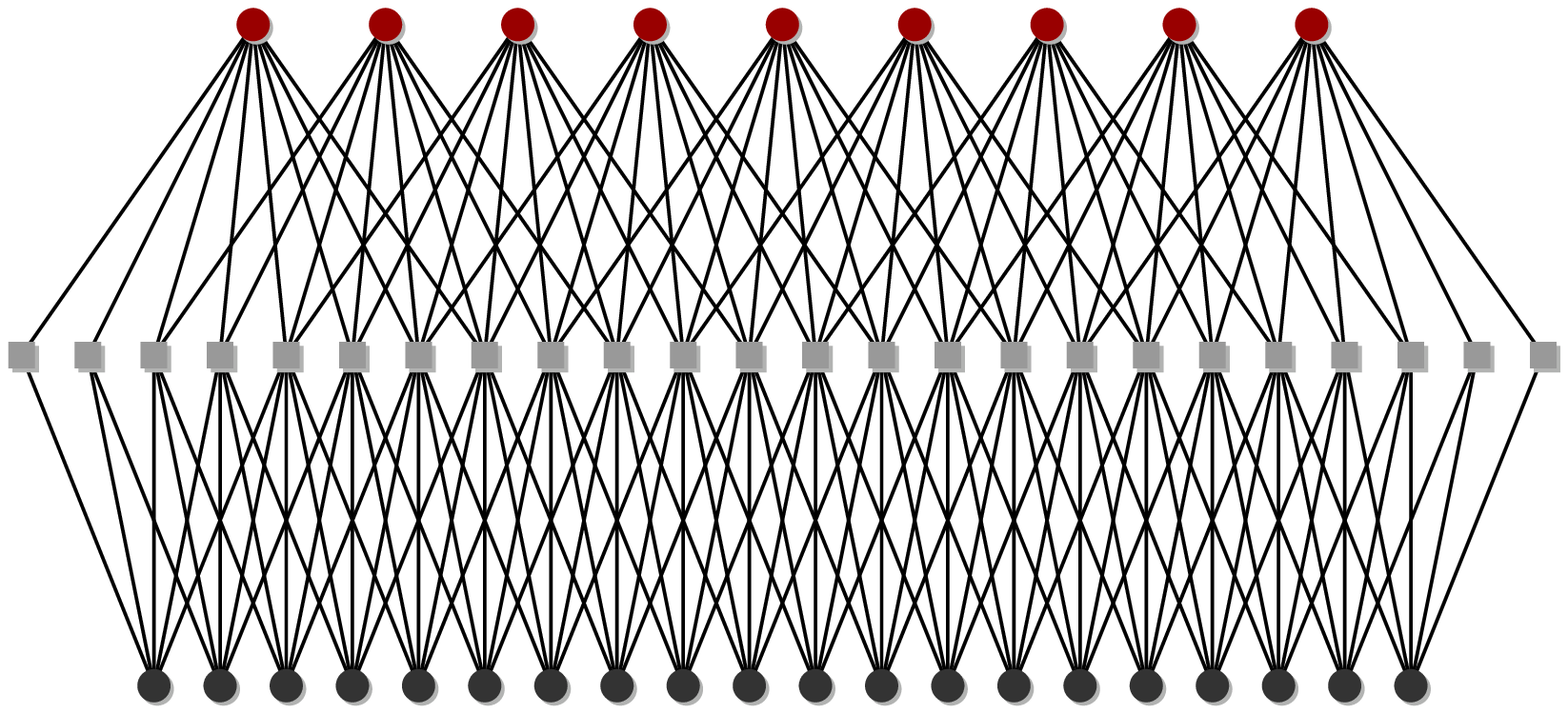}
 \end{center}
\caption{Protograph of $(8, 4, 5, 4)$ SC-MN codes. Red variable nodes are punctured. Black variable nodes are transmitted. }
\label{221857_30Sep10}
\end{figure}

In Fig.~\ref{221857_30Sep10}, we show a protograph of $(8, 4, 5, 4)$ SC-MN codes.
In the protograph of $(\dl,\dr,\dg,L)$ SC-MN codes, there are $\Vp=\hL$ punctured variable nodes 
$\Vt=\dk\hL+\dl-\dg+1$ unpunctured variable nodes, 
and $C=\dk\hL+\dl-2$ check nodes. 
Hence, in the limit of large $L$, the design coding rate is given as
\begin{align}
&R^{\mathrm{MN}}(\dl,\dr,\dg,L)=\frac{\Vp+\Vt-C}{\Vt}\nonumber\\
&=\frac{\hL+(\dk\hL+\dl-\dg+1)-(\dk\hL+\dl-2)}{\dk\hL-\dg+1}\nonumber\\
&=\frac{\hL-\dg+3}{\dk\hL-\dg+1}=\frac{1}{\dk}\quad (\hL\to\infty).\label{031351_10Jun11}
\end{align}
The density is given by
\begin{align*}
d^{\mathrm{MN}}(\dl,\dr,\dg,L)
 &=\frac{\dl\Vp+\dg\Vt}{R^{\mathrm{MN}}(\dl,\dr,\dg,L)\Vt}\\
% &=\frac{\dl\hL+\dg(\dk\hL+\dl-\dg+1)}{\hL-\dg+3}\\
 &=\dl+\dg\dk\quad (\hL\to\infty).
\end{align*}
The minimum density $4\dk=4/R^{\mathrm{MN}}(\dl,2,2,\infty)$ is attained when $\dr=\dg=2$ and $L=\infty$.

Table \ref{231057_30Sep10} shows the BP threshold values and rates of $(4,2,2,L)$ SC-MN codes. 
As increasing $L$, it is observed that  $\epsilon_{\mathrm{BP}}^{\mathrm{MN}}(4,2,2,L)$ approaches a value close to 1/2. 
From Eq.~\eqref{031351_10Jun11}, it holds that $R^{\mathrm{MN}}(4,2,2,\infty)=1/2.$
This observation supports that threshold saturation occurs by SC-MN codes. 
\begin{table}[t]
\caption{BP threshold values and design coding rate of $(4,2,2,L)$ SC-MN codes.
From Eq.~\eqref{031351_10Jun11}, it holds that $R^{\mathrm{MN}}(4,2,2,\infty)=1/2.$}
\label{231057_30Sep10}
\begin{center}
  \begin{tabular}{ccc}
 $L$& $\epsilon_{\mathrm{BP}}^{\mathrm{MN}}(4,2,2,L)$ &$R^{\mathrm{MN}}(4,2,2,L)$\\\hline
 2 &0.561146 &0.363636 \\
 4 &0.511397 &0.421053 \\
 8& 0.500252 &0.457143\\
 16& 0.499977 &0.477612\\
 32& 0.499908 &0.488550
 \end{tabular}
\end{center}
\end{table}
% L= 4 threshold=0.561146 Rate=0.363636 num_vtype=16 num_ctype=12 num_puncture=5
% L= 8 threshold=0.511397 Rate=0.421053 num_vtype=28 num_ctype=20 num_puncture=9
% L=16 threshold=0.500252 Rate=0.457143 num_vtype=52 num_ctype=36 num_puncture=17
% L=32 threshold=0.499977 Rate=0.477612 num_vtype=100 num_ctype=68 num_puncture=33
% L=64 threshold=0.499908 Rate=0.48855 num_vtype=196 num_ctype=132 num_puncture=65

%%%%%%%%%%%%%%%%%%%%%%%%%%%%%%%%%%%%%%
\subsection{Spatially-Coupled Hsu-Anastasopoulos Codes}
\begin{figure}[t]
 \begin{center}
 \includegraphics[scale=0.5]{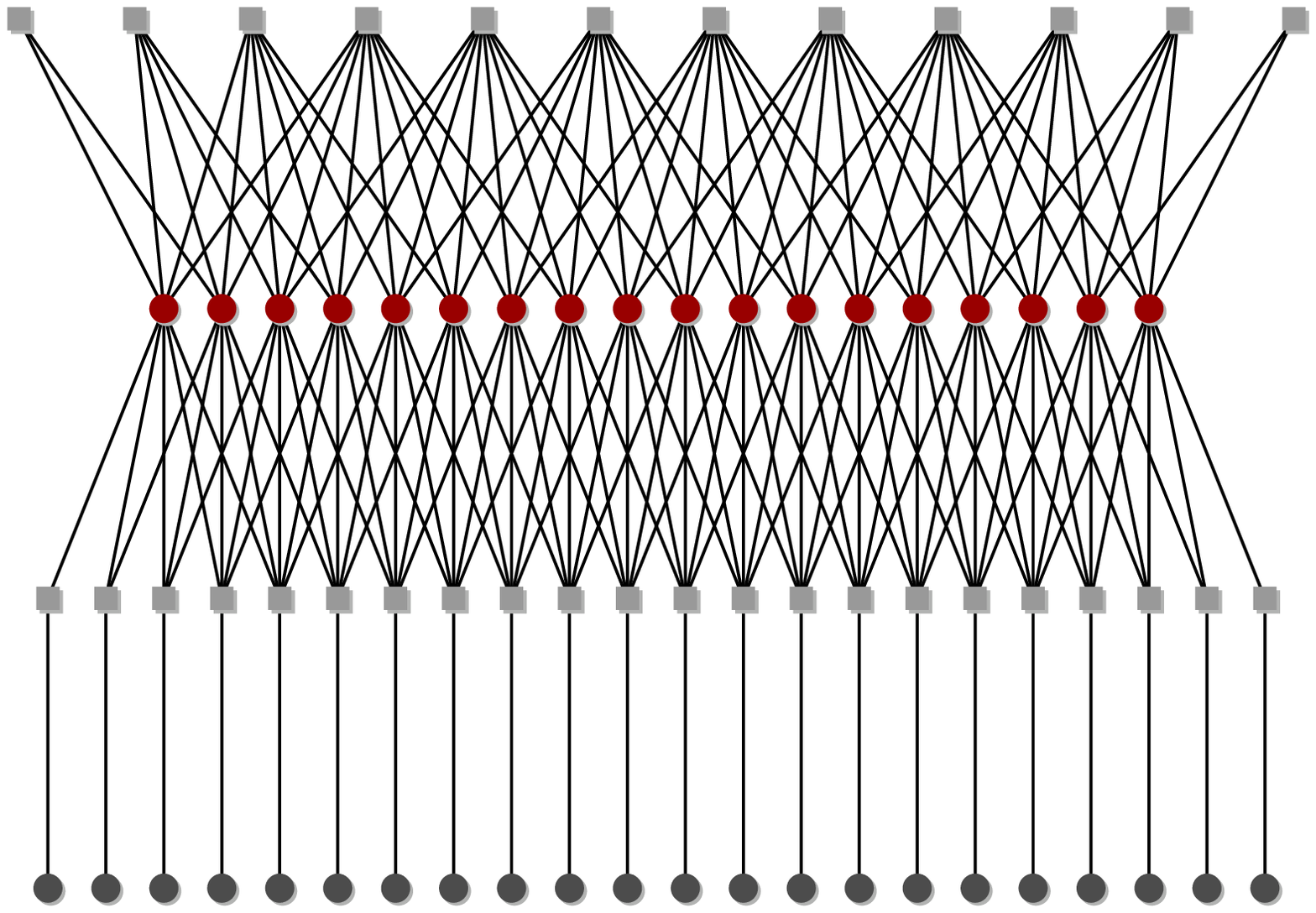}
 \end{center}
\caption{Protograph of $(8, 4, 5, 9)$ SC-HA codes. Red variable nodes are punctured. Black  variable nodes are transmitted. }
\label{163417_30Sep10}
\end{figure}

%%%%%%%%%%%%%%%%%%%%%%%%%%%%%%%%%%%%%%
In this section, we propose spatial coupling of HA codes and evaluate the BP threshold values. 
An $(\dl,\dr,\dg,L)$ SC-HA code is defined as a protograph code which is defined by base matrix
\begin{align*}
 &\begin{pmatrix}
   H(\dl,\dr,\hL,\dk)&O\\
   S(\dg,\dk\hL)&I
  \end{pmatrix}
\end{align*}
and such that the bits corresponding to the left sub-matrix are punctured.
Figure \ref{163417_30Sep10} shows the protograph $(4, 8, 5, 9)$ SC-HA codes.

We assume that $\frac{\dr}{\dl}=:\dk\in\mathbb{Z}.$
The protograph of the $(\dl,\dr,\dg,L)$ SC-HA codes has $\Vp=\dk\hL$ punctured information bit node, 
$\Vt=(\dk\hL+\dg-1)$ unpunctured bit nodes, and 
$C=(\hL+\dl-1)+(\dk\hL+\dg-1)$ check nodes. The design coding rate  is given by 
\begin{align}
&R^{\mathrm{HA}}(\dl,\dr,\dg,L)=\frac{\Vp+\Vt-C}{\Vt}\nonumber\\
&\quad=\frac{(\dk-1)\hL-\dl+1}{\dk\hL+\dg-1}=\frac{\dk-1}{\dk}\quad(\hL\to\infty).\label{031300_10Jun11}
\end{align}
The density is given by
\begin{align*}
d^{\mathrm{HA}}(\dl,\dr,\dg,L)
 &=\frac{(\dl+\dg)\Vp+\Vt}{R^{\mathrm{HA}}(\dl,\dr,\dg,L)\Vt}\\
% &=\frac{\dl\hL+\dg(\dk\hL+\dl-\dg+1)}{\hL-\dg+3}\\
 &=\frac{\dk}{\dk-1}(1+\dg+\dl)\quad (\hL\to\infty).
\end{align*}
The minimum density $\frac{5\dk}{\dk-1}=5/R^{\mathrm{HA}}(2,\dr,2,\infty)$ is attained when $\dl=\dg=2$ and $L=\infty$.

Table \ref{230737_30Sep10} shows the BP threshold values of $(2,4,2,\hL)$ SC-HA code and design coding rates.
As increasing $L$, it is observed that  $\epsilon_{\mathrm{BP}}(2,4,2,L)$ approach a value close to 1/2. 
From Eq.~\eqref{031300_10Jun11}, it holds that $R^{\mathrm{HA}}(2,4,2,\infty)=1/2.$
This observation supports that threshold saturation occurs by SC-HA codes. 
\begin{table}[t]
\caption{BP threshold values and design coding rate of $(2,4,2,L)$ SC-HA codes.
From Eq.~\eqref{031300_10Jun11}, it holds that $R^{\mathrm{HA}}(2,4,2,\infty)=1/2.$
}
\label{230737_30Sep10}
\begin{center}
  \begin{tabular}{ccc}
 $\hL$& $\epsilon_{\mathrm{BP}}^{\mathrm{HA}}(2,4,2,L)$ &$R^{\mathrm{HA}}(2,4,2,L)$\\\hline
 1&  0.695420 &0.285714\\
 2&  0.594441 &0.363636\\
 4&  0.516970 &0.421053\\
 8&  0.500460 &0.457143\\
 16& 0.499980 &0.477612\\
 32& 0.499909 &0.488550
 \end{tabular}
\end{center}
\end{table}

 {\bf Discussion}:
 We have proposed spatial-coupling of MN codes and HA codes. 
 It is observed that the BP threshold values for BEC are very close to the Shannon limit.
 In other words, threshold saturation occurs for SC-MN codes and SC-HA codes.
 We observed such threshold saturation for various parameters with $\dl,\dr,\dg \ge 2$. 
 However, these BP threshold values leave a small gap to the Shannon limit.
 Such a gap was also observed when evaluating the BP threshold of the SC-LDPC codes \cite{5695130}.
 Kudekar {\it et al.} proved that this is explained by wiggles appearing in EBP EXIT curves \cite{5695130}. 
 In order to decrease the gap, they introduced randomized SC-LDPC codes $(\dl,\dr,L,w)$ that allow connections of edges with window size $w$. 
 We observed the same effect on wiggles at the EBP EXIT curve of SC-MN codes. We will discuss the above observation in the next section.
%%%%%%%%%%%%%%%%%%%%%%%%%%%%%%%%%%%%%%%%%%%%%%%%%%%%
\section{Randomized Spatially-Coupled MacKay-Neal Codes}
%%%%%%%%%%%%%%%%%%%%%%%%%%%%%%%%%%%%%%%%%%%%%%%%%%%%
\begin{figure*}[t]
\setlength{\unitlength}{1.0bp}%
\begin{picture}(400,180)(0,0)
\put(0,0)
{
\put(0,10){\includegraphics[width=0.5\textwidth]{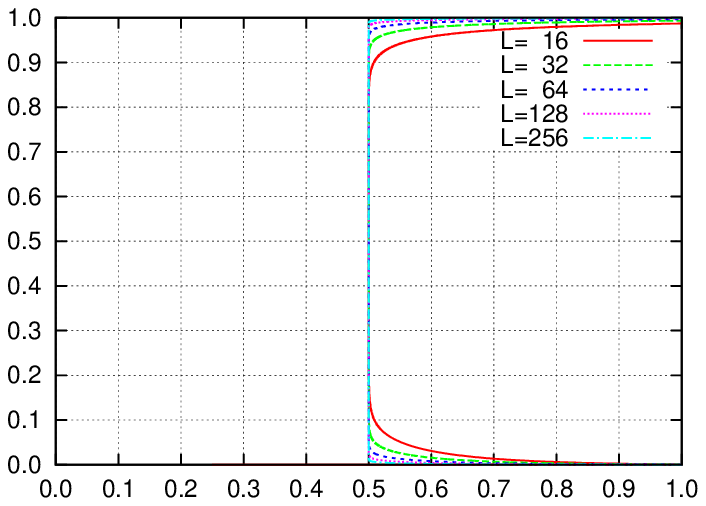} }
\put(0, 90){\rotatebox{90}{$h^{\mathrm{EBP}}(\epsilon)$}}
\put(240, 90){\rotatebox{90}{$h^{\mathrm{EBP}}(\epsilon)$}}
\put(240,10){ \includegraphics[width=0.5\textwidth]{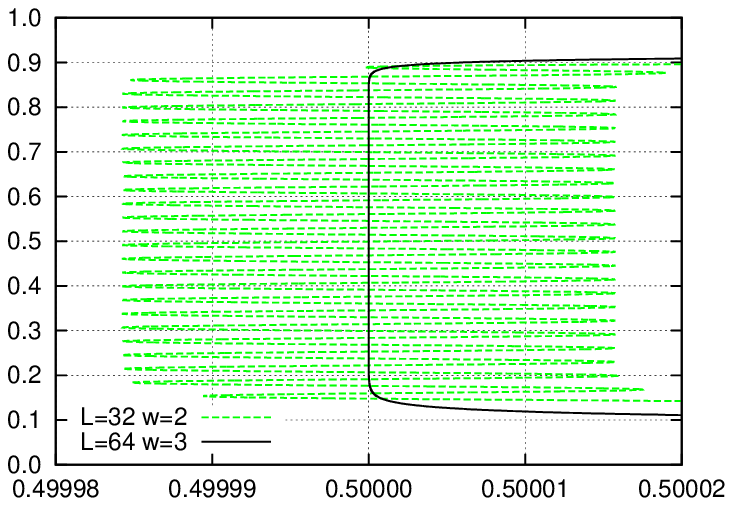} }
\put(0,0){(a)}
\put(130,10){$\epsilon$}
\put(240,0){(b)}
\put(370,10){$\epsilon$}
}
\end{picture}
\caption{
(a): The EBP EXIT curve of the $(4,2,2,L,w=2)$ SC-MN code ensemble. \\
(b): A closer look of the EBP EXIT curve of the $(4,2,2,L,w)$ SC-MN code ensemble. 
}
\label{121620_16Feb11}
\end{figure*}
In this section, we consider randomized SC-MN codes. 
Randomized SC-HA codes can be considered in a similar way. 
For simplicity, we focus only on randomized SC-MN codes.

We define an $(\dl, \dr, \dg, L, w)$ SC-MN code ensemble as follows. 
The Tanner graph of a code in the $(\dl, \dr, \dg, L, w)$ SC-MN code ensemble is constructed as follows. 
At each section $i\in\mathbf{Z}:=\{\dotsc,-2,-1,0,1,2,\dotsc\}$, consider
$\frac\dr\dl M$ information nodes of degree $\dl$, $M$ parity nodes of degree $\dg$, and $M$ check nodes which has $\dr$ incident information nodes and $\dg$ parity nodes. 
Connect randomly these nodes in such a way that 
for $i\in \mathbf{Z}$ and $j=0,\dotsc, w-1$, 
information nodes at section $i$ and check nodes at section $i+j$ are connected with $\frac{\dr M}{w}$ edges and 
parity nodes at section $i$ and check nodes at section $i+j$ are connected with $\frac{\dg M}{w}$ edges. 
Shorten the information and parity bits at section $|i|>L$, i.e., set the bits to zero and do not transmit them. 
Puncture the information nodes at section $|i|\le L$, i.e., the bits are not transmitted. 
Note that this ensemble is nicely represented by joint degree distributions \cite{Detailed_IEICE}.
The definition of the $(\dl, \dr, \dg, L, w)$ SC-MN code ensemble is based on that of $(\dl,\dr,L,w)$ randomized SC-LDPC code ensemble. 
For more details on $(\dl,\dr,L,w)$ randomized SC-LDPC code ensemble, we refer the readers to \cite[Section II.B]{5695130}.

Denote the number of transmitted bit nodes, punctured bit nodes by $\Vt$, $\Vp$, respectively.
\begin{align*}
 \Vt&=\hL M, \quad \Vp=\frac{\dr}{\dl}\hL M.
\end{align*}
The number of check nodes of degree at least 1, denoted by $C$, can be counted by the same way as 
in \cite[Lemma 3]{5695130} as follows. 
\begin{align*}
 C&=M[2L-w+2\sum_{i=0}^w(1-(\frac{i}{w})^\dr(\frac{i}{w})^{\dg})].
\end{align*}
The design coding rate $R^{\mathrm{MN}}(\dl,\dr,L,w)$ is given by 
\begin{align*}
 &R^{\mathrm{MN}}(\dl,\dr,L,w):=\frac{\Vt+\Vp-C}{V_\mathrm{t}}\\
%  &=\frac{\hL M+\frac{\dr}{\dl}\hL M-M[2L-w+2\sum_{i=0}^w(1-(\frac{i}{w})^\dr(\frac{i}{w})^{\dg})]}{\hL M}\\
%  &=\frac{(\hL)+\frac{\dr}{\dl}(\hL)-[2L-w+2\sum_{i=0}^w(1-(\frac{i}{w})^\dr(\frac{i}{w})^{\dg})]}{(\hL)}\\
% &=\frac{\dr}{\dl}+\frac{1-[-w+2\sum_{i=0}^w(1-(\frac{i}{w})^\dr(\frac{i}{w})^{\dg})]}{\hL}\\
 &=\frac{\dr}{\dl}+\frac{1+w-2\sum_{i=0}^w(1-(\frac{i}{w})^\dr(\frac{i}{w})^{\dg})}{\hL}=\frac{\dr}{\dl} \quad (\hL\to\infty).
\end{align*}

Let $x_i^{(\ell)}$ and $y_i^{(\ell)}$ be the erasure probability of messages emitting from information-bit and parity-bit nodes, respectively, at section $i$ at the $\ell$-th round of BP decoding in the limit of large $M$.
DE update equations of the randomized $(\dl, \dr, \dg, L, w)$ SC-MN code are given as follows. 
For $|i|>L$, $x_i^{(\ell)}=y_i^{(\ell)}=0$ for $\ell\ge 0$. 
For $|i|\le L$, $x_i^{(0)}=y_i^{(0)}=1$ for $\ell\ge 0$. 
For $|i|\le L$, 
\begin{align*}
&x_i^{(\ell+1)}=\\
&(\frac{1}{w}\sum_{j=0}^{w-1}[1-(1-\frac{1}{w}\sum_{k=0}^{w-1}x_{{i+j}-k}^{(\ell)})^{\dr-1}(1-\frac{1}{w}\sum_{k=0}^{w-1}y_{{i+j}-k}^{(\ell)})^{\dg}])^{\dl-1}
\end{align*}
\begin{align*}
&y_i^{(\ell+1)}=\\
&\epsilon(\frac{1}{w}\sum_{j=0}^{w-1}[1-(1-\frac{1}{w}\sum_{k=0}^{w-1}x_{{i+j}-k}^{(\ell)})^{\dr}(1-\frac{1}{w}\sum_{k=0}^{w-1}y_{{i+j}-k}^{(\ell)})^{\dg-1}])^{\dg-1}
\end{align*}
Consider fixed points of the DE system, i.e., $(\underline{x}:=(x_{-L},\dotsc,x_{L}),\underline{y}:=(y_{-L},\dotsc,y_{L}),\epsilon)$ such that $x_i=y_i=0$ for $|i|>L,$ and
{\small
 \begin{align*}
x_i=&\bigl(\frac{1}{w}\sum_{j=0}^{w-1}[1-(1-\frac{1}{w}\sum_{k=0}^{w-1}x_{{i+j}-k})^{\dr-1}(1-\frac{1}{w}\sum_{k=0}^{w-1}y_{{i+j}-k})^{\dg}]\bigr)^{\dl-1}\\
y_i=&\epsilon\bigl(\frac{1}{w}\sum_{j=0}^{w-1}[1-(1-\frac{1}{w}\sum_{k=0}^{w-1}x_{{i+j}-k})^{\dr}(1-\frac{1}{w}\sum_{k=0}^{w-1}y_{{i+j}-k})^{\dg-1}]\bigr)^{\dg-1}
\end{align*}
}
For any $\epsilon\in [0,1]$, a fixed point $(\epsilon,\underline{x}=\underline{0}, \underline{y}=\underline{0})$ is called trivial. 
Trivial fixed points is corresponds to the message density $(\underline{x},\underline{y})=(\underline{0},\underline{0})$ of successful decoding. 
The EBP EXIT curve is defined as the projected plots $(\epsilon, h^{\mathrm{EBP}}(\epsilon))$ of fixed points $(\epsilon,\underline{x},\underline{y})$, other than trivial ones, onto the following EXIT function. 
{\begin{align*}
h^{\mathrm{EBP}}(\epsilon)=&\frac{1}{\hL}\sum_{i=-L}^{L}(\frac{1}{w}\sum_{j=0}^{w-1}[1-(1-\frac{1}{w}\sum_{k=0}^{w-1}x_{{i+j}-k})^{\dr}\\
&\cdot(1-\frac{1}{w}\sum_{k=0}^{w-1}y_{{i+j}-k})^{\dg-1}])^{\dg}.
\end{align*}
}
Figure \ref{121620_16Feb11} (a) plots the EBP EXIT curve of the $(4,2,2,L,w=2)$ SC-MN code ensemble. 
Consider no points of the EBP curve $(\epsilon,h^{\mathrm{EBP}}(\epsilon))$ are at $\epsilon\in[0,\epsilon']$. 
This suggests that there are only trivial fixed points of DE with  $\epsilon\in[0,\epsilon']$ and $(\underline{x}^{(\ell)},\underline{y}^{(\ell)})$ converges to the trivial fixed point. 
It follows that the BP threshold $\epsilon_{\mathrm{BP}}^{\mathrm{MN}}(\dl,\dr,\dg,L,w)$ is given by $\epsilon$ at which the left-most cliff edge of the curve vertically drops \cite{5695130}. 
The BP threshold in Fig.~\ref{121620_16Feb11} (a) is very close to 1/2. 
However, there exists a small gap. 
A closer look at  the cliff edge of the green curve given in Fig.~\ref{121620_16Feb11} (b) reveals the cliff of $(4,2,2,L=32,w=2)$ SC-MN codes have wiggles. 
The wiggle size decreases by increasing $w$ as seen in Fig.~\ref{121620_16Feb11} (b). 

\begin{figure}[t]
\setlength{\unitlength}{1.0bp}%
\begin{picture}(200,180)(0,0)
\put(0,0)
{
\put(0,0){\includegraphics[width=0.5\textwidth,height=6.3cm]{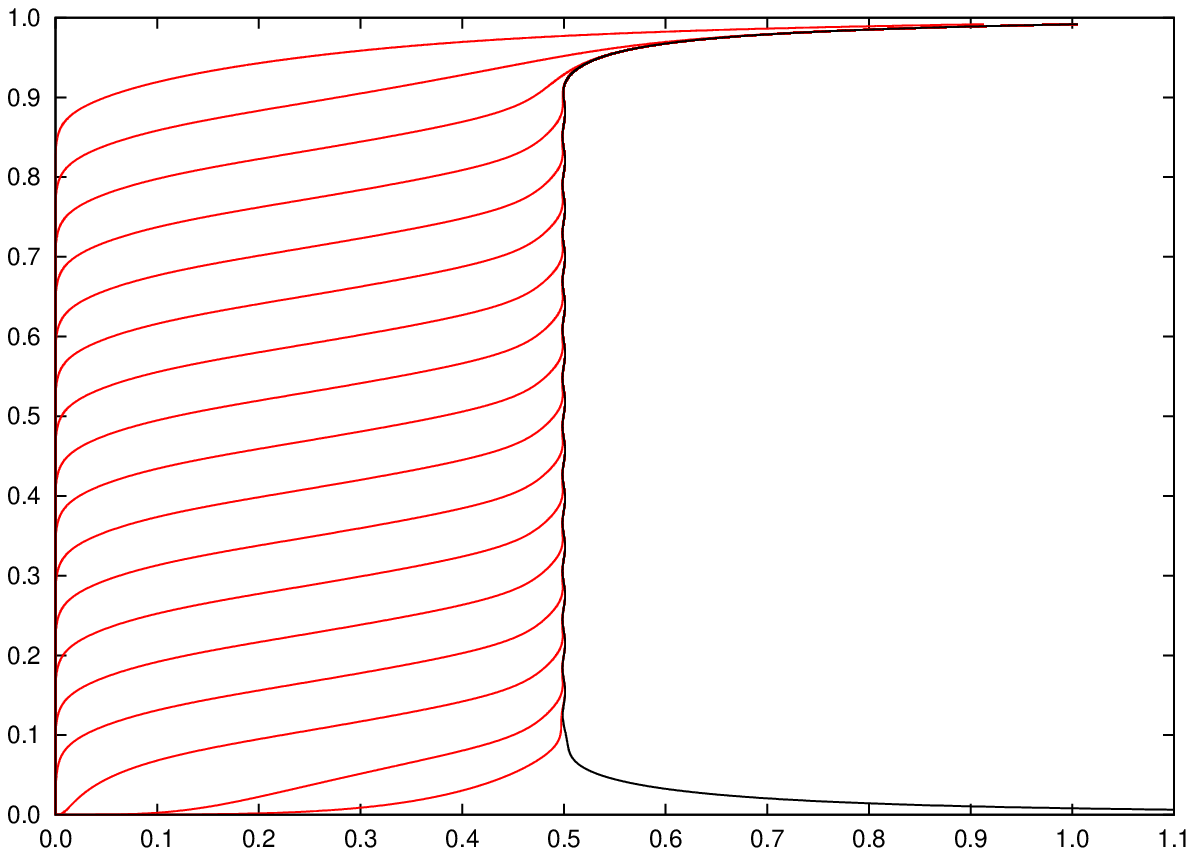}}
%\multiputlist(0,-8)(32,0)[cb]{$~$,$0.2$,$0.4$,$0.6$,$0.8$,$1.0$}
\put(30,161){\rotatebox{13}{\scriptsize $i=\pm 16$}}
\put(30,151){\rotatebox{13}{\scriptsize $i=\pm 15$}}
\put(30,141){\rotatebox{13}{\scriptsize $i=\pm 14$}}
\put(30,131){\rotatebox{13}{\scriptsize $i=\pm 13$}}
\put(30,121){\rotatebox{13}{\scriptsize $i=\pm 12$}}
\put(30,111){\rotatebox{13}{\scriptsize $i=\pm 11$}}
\put(30,101){\rotatebox{13}{\scriptsize $i=\pm 10$}}
\put(30, 92){\rotatebox{13}{\scriptsize $i=\pm 9$}}
\put(30, 81){\rotatebox{13}{\scriptsize $i=\pm 8$}}
\put(30, 71){\rotatebox{13}{\scriptsize $i=\pm 7$}}
\put(30, 61){\rotatebox{13}{\scriptsize $i=\pm 6$}}
\put(30, 51){\rotatebox{13}{\scriptsize $i=\pm 5$}}
\put(30, 42){\rotatebox{13}{\scriptsize $i=\pm 4$}}
\put(30, 32){\rotatebox{13}{\scriptsize $i=\pm 3$}}
\put(30, 21){\rotatebox{13}{\scriptsize $i=\pm 2$}}
\put(40, 11){\rotatebox{13}{\scriptsize $i=\pm 1$}}
\put(80, 12){\rotatebox{13}{\scriptsize $i=0$}}
\put(150, 25){\rotatebox{0}{$(\epsilon,h^{\mathrm{EBP}})$}}
\put(0, 90){\rotatebox{90}{$h^{\mathrm{EBP}}$}}
\put(118, -5){\rotatebox{0}{$\epsilon,y_i$}}
}
\end{picture}
\caption{
The EBP EXIT curve $(\epsilon,h^{\mathrm{EBP}})$ and $(y_i,h^{\mathrm{EBP}})$ of the $(4,2,3,L=16, w=2)$ SC-MN code ensemble. 
}
\label{121038_16Feb11}
\end{figure}

\begin{figure}[t]
\setlength{\unitlength}{1.0bp}%
\begin{picture}(200,180)(0,0)
\put(0,0)
{
\put(0,0){\includegraphics[width=0.5\textwidth,height=6.3cm]{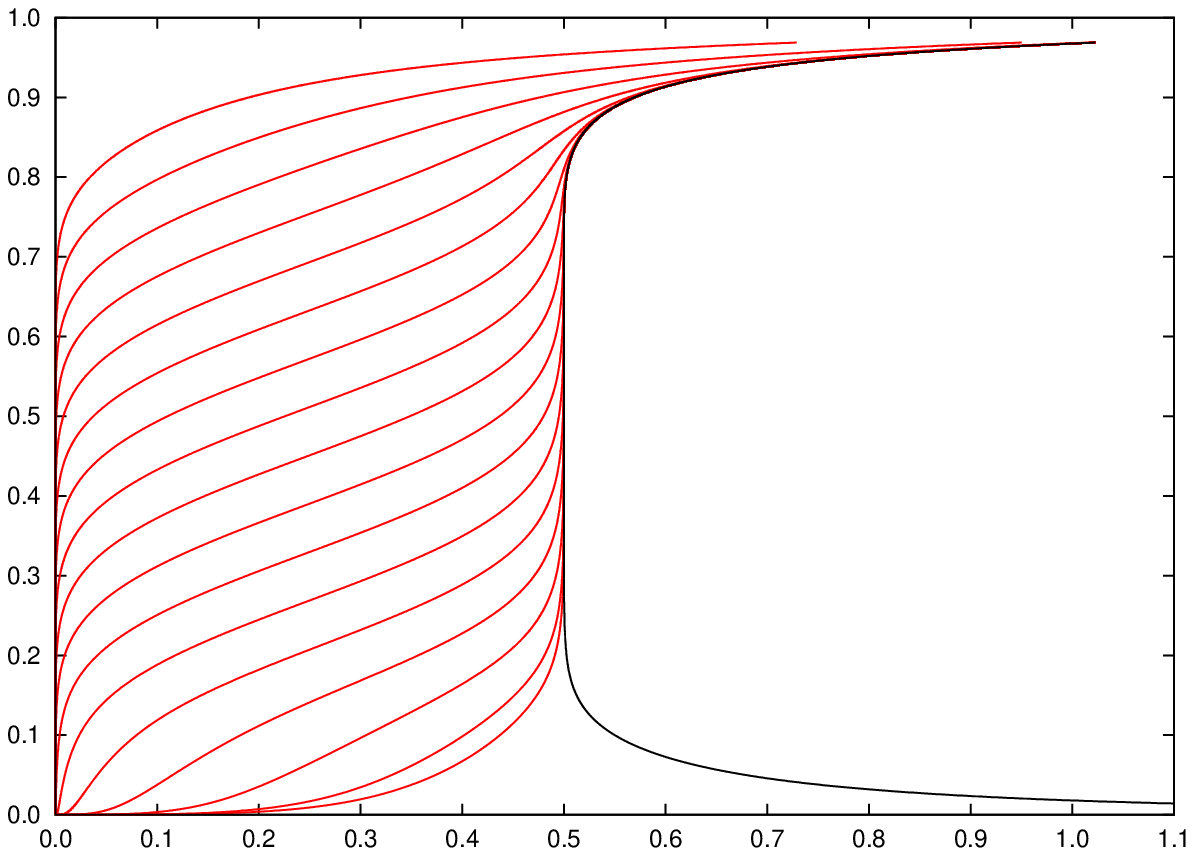}}
%\multiputlist(0,-8)(32,0)[cb]{$~$,$0.2$,$0.4$,$0.6$,$0.8$,$1.0$}
\put(35,155){\rotatebox{15}{\scriptsize $i={\pm 16}$}}
\put(35,143){\rotatebox{20}{\scriptsize $i={\pm 15}$}}
\put(35,133){\rotatebox{20}{\scriptsize $i={\pm 14}$}}
\put(35,123){\rotatebox{20}{\scriptsize $i={\pm 13}$}}
\put(35,113){\rotatebox{20}{\scriptsize $i={\pm 12}$}}
\put(35,103){\rotatebox{21}{\scriptsize $i={\pm 11}$}}
\put(35, 93){\rotatebox{22}{\scriptsize $i={\pm 10}$}}
\put(35, 83){\rotatebox{22}{\scriptsize $i={\pm 9}$}}
\put(35, 73){\rotatebox{22}{\scriptsize $i={\pm 8}$}}
\put(35, 63){\rotatebox{22}{\scriptsize $i={\pm 7}$}}
\put(35, 53){\rotatebox{22}{\scriptsize $i={\pm 6}$}}
\put(35, 43){\rotatebox{22}{\scriptsize $i={\pm 5}$}}
\put(35, 33){\rotatebox{22}{\scriptsize $i={\pm 4}$}}
\put(40, 23){\rotatebox{22}{\scriptsize $i={\pm 3}$}}
\put(60, 22){\rotatebox{22}{\scriptsize $i={\pm 2}$}}
\put(80, 20){\rotatebox{24}{\scriptsize $i={\pm 1}$}}
\put(100, 13){\rotatebox{30}{\scriptsize $i={0}$}}
\put(150, 25){\rotatebox{0}{$(\epsilon,h^{\mathrm{EBP}})$}}
\put(0, 90){\rotatebox{90}{$h^{\mathrm{EBP}}$}}
\put(118, -5){\rotatebox{0}{$\epsilon,y_i$}}
}
\end{picture}
\caption{
The EBP EXIT curve $(\epsilon,h^{\mathrm{EBP}})$ and $(y_i,h^{\mathrm{EBP}})$ of the $(4,2,3,L=16,w=4)$ SC-MN code ensemble. 
}
\label{121054_16Feb11}
\end{figure}

Figure \ref{121038_16Feb11} and \ref{121054_16Feb11} plot
the EBP EXIT curves $(\epsilon,h^{\mathrm{EBP}})$ and $(y_i,h^{\mathrm{EBP}})$ of the $(4,2,3,16, w)$ SC-MN code ensemble for $w=2$ and $w=4$, respectively. 
It is observed that as decreasing $h^{\mathrm{EBP}}$ value, $y_i$ starts to collapse from the boundaries $|i|=L$ and gradually to the center $i=0$. 
It is observed that uncollapsed $y_i$ takes almost the same value $0.5$. 
Each time $y_i$ collapse, $\epsilon$ is affected a little, which gives rise to a wiggle. 
By increasing $w$, each $y_i$ collapses slowly, which gives rise to a smaller wiggle. 
%%%%%%%%%%%%%%%%%%%%%%%%%%%%%
\section{Conclusion}
%%%%%%%%%%%%%%%%%%%%%%%%%%%%%
 We have proposed spatial-coupling of MN (resp.~HA) codes with bounded density.
 By DE analysis, we observed that the BP threshold values of the SC-MN (resp.~SC-HA) codes for BEC are very close to the Shannon limit.
       This empirical evidence supports that the SC-MN (resp.~SC-HA) codes may achieve the Shannon limit of the BEC.
 In other words, threshold saturation occurs for SC-MN codes and SC-HA codes.
 We observed such threshold saturation for various parameters with $\dl,\dr,\dg \ge 2$.

 We further investigate the EBP EXIT curve of randomized SC-MN codes.
 We observed that the small gap between BP threshold and the Shannon limit is caused by wiggles of BP EXIT curve.
 By increasing the randomized window size $w$, the wiggle size largely decreased.
%%%%%%%%%%%%%%%%%%%%%%%%%%%%%
% \begin{table}[t]
% \caption{$B#285>C<:DL?.O)$K$*$1$k(B$(3,6,3,L)$-$B6u4V7k9g(BHA$BId9f$N(BBP$BogCM(B$\epsilon_{\mathrm{BP}}$$B$H(B
% $B@_7WId9f2=N((B$R$}
% \label{230737_30Sep10}
% \begin{center}
%   \begin{tabular}{ccc}
%  $\hL$& $\epsilon_{\mathrm{BP}}\approx$ &$R$\\\hline
% 1  &0.707656&0.125000 \\
% 2   &0.545744&0.250000\\
% 3  &0.504719&0.312500 \\
% 4   &0.499718&0.350000\\
% 8  &0.499411&0.416667 \\
% 16 &0.499362&0.455882 \\
% 32  &0.499250&0.477273   
%  \end{tabular}
% \end{center}
% \end{table}
%%%%%%%%%%%%%%%%%%%%%%%%%%%%%%
%\appendix{proof}
%%%%%%%%%%%%%%%%%%%%%%%%%%%%%%
%\input{proof43012984}
%%%%%%%%%%%%%%%%%%%%%%%%%%%%%%%%%
%%%%%%%%%%%%%%%%%%%%%%%%%%%%%%%%%
%\section{Conclusion}
%%%%%%%%%%%%%%%%%%%%%%%%%%%%%%%%%
%%%%%%%%%%%%%%%%%%%%%%%%%%%%%%%%%
\section*{Acknowledgment}
The first author would like to thank, chronologically, Tadashi Wadayama, Michael Lentmaier, R\"udiger Urbanke, David Saad, Ryuhei Mori, Kazushi Mimura, Hironori Uchikawa, Shrinivas Kudekar,
Toshiyuki Tanaka, and Yoshiyuki Kabashima for their help and useful discussions.
He is also grateful to the associate editor and the anonymous reviewers for their helpful and constructive comments.
%%%%%%%%%%%%%%%%%%%%%%%%%%%%%%%%%%
%%%%%%%%%%%%%%%%%%%%%%%%%%%%%%%%%
\bibliographystyle{ieicetr}
\bibliography{IEEEabrv.bib,../../../kenta_bib}
\end{document}